\documentclass[aps,prl,reprint,superscriptaddress]{revtex4-2}
\usepackage{bibunits}
\usepackage{hyperref}

\usepackage{etoolbox}
\newbool{isdraft}
\setbool{isdraft}{false} 

\ifbool{isdraft}{
    \newcommand{\dfsection}[1]{%
        \par\textcolor{blue!80!black}{\emph{#1.}---}\ignorespaces%
    }
}{
    \newcommand{\dfsection}[1]{%
        \par\emph{#1.}---\ignorespaces%
    }
}

\usepackage[utf8]{inputenc}

\usepackage{soul,xcolor}
\usepackage{graphicx}
\usepackage{amsmath}
\usepackage{amssymb}
\usepackage{verbatim}
\usepackage{mathrsfs}
\usepackage{array}
\usepackage{layout}
\usepackage{textcomp}
\usepackage{latexsym}
\usepackage{slashed}
\usepackage{rotating}
\usepackage{booktabs}
\usepackage{relsize}
\usepackage{multirow}
\usepackage{bigints}
\usepackage{physics}

\usepackage{cleveref}
\crefname{figure}{Fig.}{Figs.}
\crefname{section}{Sec.}{Secs.}
\Crefname{figure}{Figure}{Figures}
\crefname{equation}{Eq.}{Eqs.}
\Crefname{equation}{Equation}{Equations}

\newcommand\cyansout{\bgroup\markoverwith{\textcolor{cyan}{\rule[0.5ex]{2pt}{0.4pt}}}\ULon}
\newcommand\blueout{\bgroup\markoverwith{\textcolor{blue}{\rule[0.5ex]{2pt}{0.4pt}}}\ULon}

\newcommand{\keV}{{\rm ~keV}}
\newcommand{\MeV}{{\rm ~MeV}}
\newcommand{\GeV}{{\rm ~GeV}}
\newcommand{\TeV}{{\rm ~TeV}}

\newcommand{\calL}{{\mathcal{L}}}
\newcommand{\calO}{{\mathcal{O}}}
\newcommand{\calM}{{\mathcal{M}}}

\newcommand{\zp}{Z'}

\newcommand{\mzp}{m_{Z'}}
\newcommand{\mchi}{m_\chi}

\newcommand{\mphi}{m_\phi}

\newcommand{\sa}{s_\alpha}
\newcommand{\tchi}{{T_\chi}}
\newcommand{\tsm}{{T_\SM}}
\newcommand{\tkd}{{T_{\rm kd}}}

\newcommand{\vmin}{v_{\rm min}}
\newcommand{\er}{E_R}
\newcommand{\SM}{{\rm SM}}
\newcommand{\roi}{\rm ROI}

\makeatletter
\def\NAT@bibsetnum#1{\setlength{\topsep}{\z@}\NATx@bibsetnum{0}}
\makeatother
\begin{document}
\begin{bibunit}[apsrev4-2]


\title{Exothermic Dark Matter with a Dark Higgs:\\ The LZ 248 keV Event and the Galactic Center Excess}


\author{Jinheung Kim}
\email{jhkim1216@kias.re.kr}
\affiliation{School of Physics, Korea Institute for
Advanced Study, Seoul 02455, Korea}

\author{Jongkuk Kim}
\email[Corresponding author: ]{jongkuk.kim927@gmail.com}
\affiliation{Division of Liberal Studies,
Kangwon National University, Samcheok 25913, Korea}

\author{Pyungwon Ko}
\email{pko@kias.re.kr}
\affiliation{Quantum Universe Center, KIAS,
Seoul 02455, Korea}




\begin{abstract}
LUX-ZEPLIN (LZ) reported a $248~\mathrm{keV}$ nuclear-recoil candidate.
Exothermic dark matter can explain it in a dark $U(1)$ model whose dark Higgs generates the dark photon mass and the mass splitting.
Since the dark photon mass sets both the excited-state fraction and the scattering rate, two dark photon masses explain the event, near $1$ and $25~\mathrm{GeV}$ for kinetic mixing $\epsilon=10^{-4}$.
Cosmic microwave background limits exclude most of the lighter branch.
On the heavier branch, the southern-sky Galactic Center $\gamma$-ray excess complements the LZ event by selecting a dark matter mass of $40$--$44~\mathrm{GeV}$, and its fit favors a dark Higgs boson near the dark photon mass.
\end{abstract}

\maketitle



\dfsection{\label{sec:intro}Introduction}
Standard direct-detection analyses of dark matter (DM)~\cite{Bertone:2004pz,Billard:2021uyg} assume spin-independent elastic scattering with a velocity-independent cross section, whose recoil spectrum falls steeply, and focus on recoil energies of $\mathcal{O}(10)~\mathrm{keV}$~\cite{LZ:2024zvo,XENON:2025vwd}.
The LUX-ZEPLIN (LZ) collaboration recently extended its nuclear-recoil search well beyond this range and reported a single candidate event at $\er \simeq 248\keV$, with a global significance of $2.6\sigma$~\cite{LZ:2026axp}.
Although a background origin cannot be excluded, the event still motivates DM explanations~\cite{Lou:2026idn,CentellesChulia:2026ats}, which must not overpredict the rates at lower and higher recoil energies.

Inelastic DM, which scatters into a heavier (endothermic) or lighter (exothermic) state, achieves this by restricting nuclear recoils to a finite energy window.
Endothermic explanations, mediated by the $Z$ boson as in the Higgsino~\cite{Freese:2026sga, Wu:2026nhi, Fan:2026kxx, Yin:2026jnn, Freese:2026mfi, Du:2026guj, Frolovsky:2026tvq, Khan:2026zuj, Khan:2026pgu, Hor:2026voa, Langhoff:2026ujr, Unwin:2026qep, Pallis:2026jdc, Jeong:2026tlf, Kotlarski:2026pep, Chattopadhyay:2026ryw, Bisal:2026khf, Lian:2026bze}, inert-doublet~\cite{Murayama:2026apt, Wang:2026ytg, Nomura:2026qyq, Okada:2026upm, Borah:2026ris}, and other electroweak-multiplet~\cite{Smirnov:2026aqk, Ahmed:2026qjg, Girmohanta:2026qzo, Paul:2026okh, Lee:2026jxl, Nagata:2026pbj, Visinelli:2026kgt, Bandyopadhyay:2026gjw, Borah:2026zwf, Lian:2026hpm, Langhoff:2026qgo, Lee:2026tro} scenarios, or by a dark photon~\cite{Zhu:2026dag, Dutra:2026fng, Das:2026buc, Barman:2026omh, Chakraborti:2026gmu, Lee:2026wof, Qi:2026vyp, Du:2026lpa, Cabo-Almeida:2026uqw, Su:2026rwz, DiMauro:2026ldr, Pradhan:2026inq, Okada:2026fef, Okada:2026eol, Kumar:2026lgi, Lee:2026xxh, Ahmed:2026com, Sheng:2026tqt, Sannino:2026hkc, Chanda:2026wdl, Sahoo:2026dnw}, among others~\cite{An:2026pkc, Yuan:2026djt, Yamashita:2026ump, De:2026win, Das:2026uyy, Asadi:2026iot, He:2026hqz, McCabe:2026crm, Mahapatra:2026glu, Choudhury:2026qjz}, typically require TeV-scale DM with a splitting of a few hundred keV and can depend strongly on the high-speed tail of the DM velocity distribution~\cite{Fan:2026kxx, Wu:2026nhi, Freese:2026mfi, Langhoff:2026ujr, Smirnov:2026aqk, Asadi:2026iot, Langhoff:2026qgo, Chanda:2026wdl, He:2026hqz, Luengas:2026tlf}.
In exothermic scattering~\cite{Graham:2010ca}, the transition to the lighter state releases energy, allowing even slow DM to produce high-energy nuclear recoils.
Explaining the LZ event in this way requires a sufficient abundance of the heavier state today~\cite{deLima:2026shq, Baer:2026fpy, Tang:2026lgu, Sheikh:2026lxx, Xing:2026civ, Jodlowski:2026wjx, Yang:2026wpb, Gemmell:2026yaw, He:2026idw, Oliveira:2026jtn, Jho:2026zzj, Dent:2026bji, Fan:2026hzw, Palmisano:2026kuj, Deka:2026zeu}.
However, the interaction that mediates nuclear scattering can also deplete the heavier state through conversion and decay, tying its present-day abundance to the scattering rate~\cite{Batell:2009vb,Finkbeiner:2009mi}.

Among the exothermic models, Refs.~\cite{deLima:2026shq,Baer:2026fpy,Deka:2026zeu} use an inelastic dark photon and Ref.~\cite{Gemmell:2026yaw} a hadrophilic $U(1)_B$ boson. The present-day excited-state fraction $f_2$ is set close to $1/2$~\cite{deLima:2026shq,Gemmell:2026yaw,Deka:2026zeu} or evolved with the conversion alone~\cite{deLima:2026shq,Baer:2026fpy}, and the mass splitting is a free parameter~\cite{deLima:2026shq,Baer:2026fpy,Deka:2026zeu} or comes from a scalar outside the relic-density and $\gamma$-ray calculations~\cite{Gemmell:2026yaw}.
Reference~\cite{Gemmell:2026yaw} also explained the LZ event together with the Galactic Center $\gamma$-ray excess (GCE; see also Ref.~\cite{Yang:2026wpb}).

In this work, DM is a Dirac fermion $\chi$ charged under a dark $U(1)_X$.
A dark Higgs field breaks this symmetry, gives the dark photon ($Z'$) its mass, and splits $\chi$ into two Majorana states, $\chi_1$ and $\chi_2$, separated in mass by $\delta$, in a gauge-invariant and theoretically consistent way~\cite{Ko:2019wxq,Jeong:2026dzt}.
We fix the dark gauge coupling to reproduce the relic density and evolve $f_2$ through the conversion $\chi_2\chi_2 \to \chi_1\chi_1$, the $\chi_2\ell \leftrightarrow \chi_1\ell$ scattering on the thermal bath, and the $\chi_2$ decays.
Both $f_2$ and the scattering cross section are then determined by $(\mchi,\delta,\mzp)$ for a given $\epsilon$.
Their interplay gives two branches that can explain the LZ event, a low-$\mzp$ branch with a large cross section but a small $f_2$ and a high-$\mzp$ branch with the reverse.
On each branch, the LZ event alone also leaves the DM mass open.

The southern-sky GCE~\cite{Cholis:2021rpp} breaks both degeneracies, selecting one branch and the DM mass.
On the low-$\mzp$ branch, the predicted $\gamma$-ray spectrum fits the GCE worse than no signal, and cosmic microwave background (CMB) limits on the $\chi_2$ decays~\cite{Slatyer:2016qyl} also exclude this branch near its likelihood minimum.
On the high-$\mzp$ branch, the LZ best fit predicts a compatible $\gamma$-ray spectrum without any fit to the GCE.
The GCE then selects $\mchi \approx 40$--$44\GeV$ and favors a dark Higgs boson near the dark photon mass through the coannihilation $\chi_1\chi_2 \to Z'\phi$.

\begin{figure*}[!th]
    \centering
    \includegraphics[width=\textwidth]{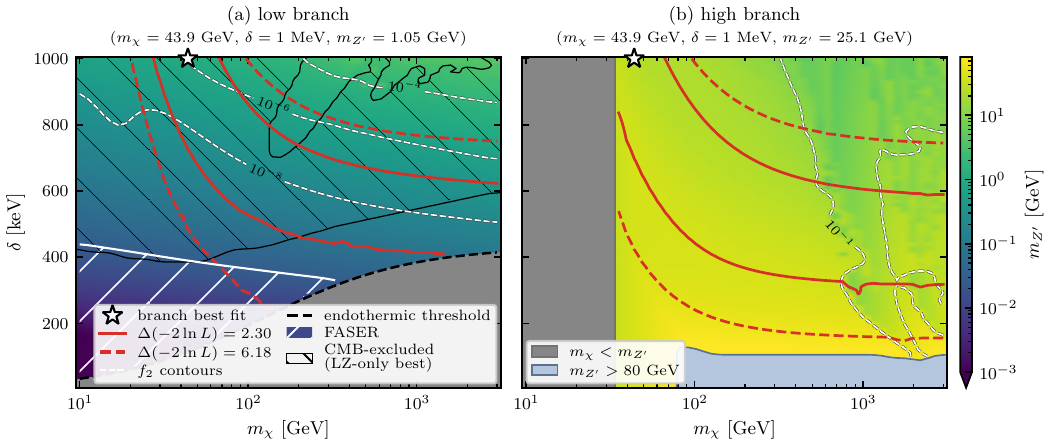}
\caption{
LZ best-fit $\mzp$ (color) and the resulting $f_2$ for the (a) low and (b) high branches, with $\epsilon = 10^{-4}$ and $\mphi = \mzp$.
Solid (dashed) red lines are the nominal $1\sigma$ ($2\sigma$) contours around the minimum of each branch (star).
Hatching marks where this best fit violates the CMB or the preliminary FASER~\cite{faser_collaboration_2026_fwshk-fak94} limit.
Gray in (a) lies below the endothermic threshold. In (b), the minimum reaches the upper end of the $\mzp$ scan in the gray and blue regions.\label{fig:allowed}}
\end{figure*}

\dfsection{Model\label{sec:model}}
We consider a dark $U(1)_X$ gauge symmetry with gauge boson $X_\mu$ and coupling $g_X$, a Dirac fermion $\chi$ with charge $Q_\chi = 1$, and a complex dark Higgs field $\Phi$ with $Q_\Phi = 2$, with the renormalizable Lagrangian~\cite{Baek:2020owl}
\begin{align}
\mathcal{L}
=&\,
\mathcal{L}_{\rm SM}
-\frac{1}{4}X_{\mu\nu}X^{\mu\nu}
-\frac{\sin\epsilon}{2}
 X_{\mu\nu}B^{\mu\nu} +\bar{\chi}(i\slashed{D} - m_D)\chi
\nonumber\\
& 
-\frac{1}{\sqrt{2}}\left(y\,\Phi^\dagger \overline{\chi^c}\chi + \mathrm{H.c.}\right)
+(D_\mu\Phi)^\dagger(D^\mu\Phi)
+m_\Phi^2\Phi^\dagger\Phi
\nonumber\\
& 
-\lambda_\Phi(\Phi^\dagger\Phi)^2
-\lambda_{H\Phi}(H^\dagger H)(\Phi^\dagger\Phi),
\label{eq:lagrangian}
\end{align}
where $\epsilon$ parametrizes the kinetic mixing with the hypercharge gauge field $B_\mu$, $X_{\mu\nu}$ and $B_{\mu\nu}$ are the field-strength tensors, $m_D$ is the Dirac mass, and $H$ is the Standard Model (SM) Higgs doublet.

In the unitary gauge, $\Phi = (v_\Phi + \phi_0)/\sqrt{2}$, and $X_\mu$ acquires the mass $m_X = 2g_Xv_\Phi$.
Through the kinetic mixing, $X_\mu$ mixes with the SM neutral gauge bosons into the dark photon $Z'$, with $\mzp \simeq m_X$.
For $\mzp \ll m_Z$, the $Z'$ couples to the electromagnetic current with strength $\epsilon c_W e$, where $c_W$ ($s_W$) is the cosine (sine) of the Weinberg angle.
We consider $\mzp \lesssim 80\GeV$, away from the $Z$ pole.

The Yukawa term splits $\chi = (\chi_2 + i\chi_1)/\sqrt{2}$ into Majorana states with masses $m_{\chi_{1,2}} = m_D \mp yv_\Phi$, i.e., $\delta = 2yv_\Phi$. We denote $m_\chi \equiv m_{\chi_1}$.
For a common Yukawa coupling of the two Weyl components of $\chi$~\cite{Baek:2020owl}, the $Z'$ couples to $\chi_{1,2}$ only off-diagonally, so that the leading, tree-level $Z'$ exchange with nuclei is inelastic, and the remnant $Z_2$ symmetry, under which only $\chi$ is odd, allows $\chi_2$ to decay only into $\chi_1$ plus $Z_2$-even particles.
The relevant interactions are
\begin{align}
    \calL \supset& -i\frac{g_X}{2}\left(Z'_\mu + \epsilon s_W Z_\mu\right)\left(\bar{\chi}_2 \gamma^\mu \chi_1 - \bar{\chi}_1\gamma^\mu\chi_2\right)
    \nonumber\\
    &-\frac{g_X\delta}{2\mzp}\,\phi\left(\bar{\chi}_2\chi_2-\bar{\chi}_1\chi_1\right).
\end{align}

The Higgs portal mixes $\phi_0$ with the SM Higgs boson into the mass eigenstates $\phi$ and $h$, with a mixing angle $\alpha$, $\sa \equiv \sin\alpha = 10^{-6}$, far below the current limits of $2.5\times10^{-4}$ at $\mphi \simeq 1\GeV$~\cite{Winkler:2018qyg,Ferber:2023iso} and $0.14$ at $25\GeV$~\cite{LEPWorkingGroupforHiggsbosonsearches:2003ing}.
The couplings of $\phi$ to $\chi_{1,2}$ are suppressed by $\delta/\mzp$, so that $\phi$ exchange contributes negligibly to the LZ signal.
The coannihilation $\chi_1\chi_2 \to Z'\phi$, in contrast, proceeds through the unsuppressed gauge coupling $\phi Z'Z'$, so $\phi$ enters both the relic density and the $\gamma$-ray spectrum.
We set $m_\phi = \mzp$ unless stated otherwise.
The dark Higgs field also makes the model consistent, generating $\delta$ gauge-invariantly and canceling the high-energy growth of amplitudes with longitudinal $Z'$ bosons~\cite{Ko:2019wxq,Jeong:2026dzt} (\cref{ssec:model}).

\dfsection{Relic density and $f_2$}
For the relic density, we set $g_X$ at each parameter point so that $\chi_{1,2}$ account for all of the DM, $\Omega_\chi h^2 = 0.120$~\cite{Planck:2018vyg}.
We compute it with micrOMEGAs~\cite{Belanger:2026asz} from $\chi_i\chi_i \to Z'Z'$ ($i = 1,2$) and $\chi_1\chi_2 \to Z'\phi$, taking $\mzp < \mchi$ so that both are open for $\mphi = \mzp$.
After chemical freeze-out, the excited-state fraction $f_2 \equiv n_{\chi_2}/(n_{\chi_1}+n_{\chi_2}) = 1 - f_1$ evolves through conversion $\chi_2\chi_2 \leftrightarrow \chi_1\chi_1$, scattering $\chi_2\ell \leftrightarrow \chi_1\ell$ with charged leptons in the thermal bath, and decays $\chi_2 \to \chi_1 + \SM$ (\cref{ssec:history}).
Starting from $f_2 \simeq 1/2$, conversions can reduce $\chi_2$ once the DM temperature $T_\chi$, obtained from kinetic decoupling, drops below $\delta$, until they freeze out.
We compute the conversion nonperturbatively~\cite{Ko:2019wxq}, as Ref.~\cite{Baer:2026fpy} did for the LZ event.
The decays of $\chi_2$ then multiply $f_2$ by $\exp(-t_0/\tau_{\chi_2})$, with $t_0$ the age of the Universe.
We restrict ourselves to $\delta \le 1\MeV$, just below $2m_e$, avoiding the rapid decay $\chi_2 \to \chi_1 e^+e^-$, so that $\chi_2$ decays only into $\chi_1\nu\bar\nu$ and $\chi_1 + 3\gamma$, with widths that scale as $g_X^2\epsilon^2/\mzp^4$ and grow steeply with $\delta$~\cite{Pospelov:2008jk,Batell:2009vb,McDermott:2017qcg}.

\begin{figure*}[!t]
\centering
\includegraphics[width=0.49\textwidth]{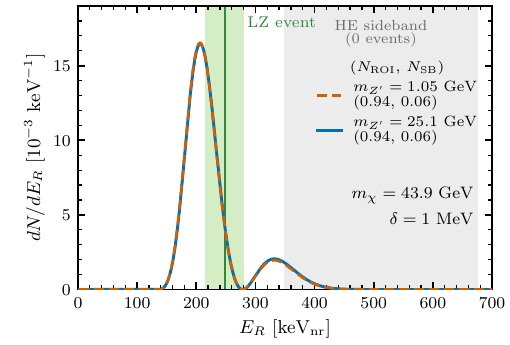}
\hfill
\includegraphics[width=0.49\textwidth]{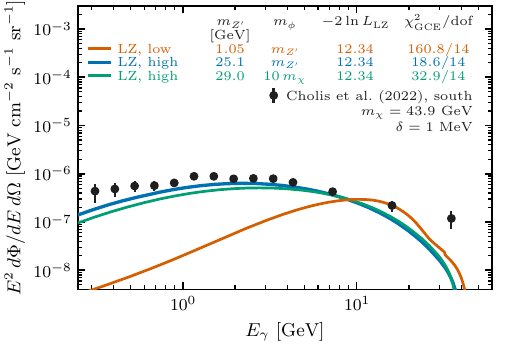}
\caption{
Left: LZ recoil spectra of the best fits on the low (orange) and high (blue) branches, before efficiency and smearing. The two nearly coincide. The green band marks the observed event ($\pm 32.5\keV$).
Right: $\gamma$-ray flux and the southern-sky GCE spectrum~\cite{Cholis:2021rpp}, with the high branch also for $\mphi = 10\,\mchi$ (green).}
\label{fig:lz-best-fit}
\end{figure*}

\dfsection{Implications for the LZ event}
In 2.84~tonne-years, the extended LZ search~\cite{LZ:2026axp} found 1710 events in its region of interest (ROI), which extends to about $270\keV$.
One of them stands out in the nuclear-recoil (NR) band, at $\er = 248 \pm 23_{\rm stat} \pm 23_{\rm sys}\keV$.
No event lies in the high-energy sideband (SB) of the same science sample, at $349 \lesssim \er \lesssim 676\keV$ in our reconstruction of the energy scale.

For the exothermic process $\chi_2 N \to \chi_1 N$, a recoil energy $\er$ requires a minimum DM velocity~\cite{Barello:2014uda,Tucker-Smith:2001myb}
\begin{equation}
    \vmin(\er) = \frac{1}{\sqrt{2 m_N \er}}\abs{\frac{m_N\er}{\mu_N}-\delta},
\label{eq:vmin}
\end{equation}
where $m_N$ is the target nucleus mass and $\mu_N$ the DM--nucleus reduced mass.
It vanishes at the kinematic reference energy $\er^* = \mu_N\delta/m_N$.
For $\er$ far from $\er^*$, the required velocity exceeds the maximum Earth-frame DM speed of the Standard Halo Model~\cite{Baxter:2021pqo}. The recoil spectrum is therefore confined to a window around $\er^*$, without overpredicting the rates at lower and higher $\er$.

The event rate is
\begin{align}
\dv{N}{\er} =&\, \mathcal{E}\,\frac{\rho_\odot}{m_\chi}\,\frac{Z^2\sigma_{\chi p}}{2\mu_{\chi p}^2}\,F^2(q^2)\left(\frac{\mzp^2}{q^2+\mzp^2}\right)^2
\nonumber\\
&\times \left[f_1\, \eta\!\left(\vmin^{\rm endo}\right) + f_2\, \eta\!\left(\vmin^{\rm exo}\right)\right],
\end{align}
where $\mathcal{E}$ is the exposure, $\rho_\odot = 0.4\GeV/{\rm cm}^3$ the local DM density, $\sigma_{\chi p} = 4\alpha_{\rm EM}\epsilon^2 c_W^2 g_X^2 \mu_{\chi p}^2/\mzp^4$ the $Z'$-mediated DM--proton cross section at zero momentum transfer, with $\mu_{\chi p}$ the DM--proton reduced mass, $F$ the Helm form factor~\cite{Helm:1956zz,Lewin:1995rx}, and $q^2 = 2m_N\er$.
The velocity integral is $\eta(\vmin) = \int_{v>\vmin} d^3v\, f(\vec v)/v$, with $f(\vec v)$ the DM velocity distribution in the Earth frame, $\vmin^{\rm exo}$ given by Eq.~\eqref{eq:vmin}, and $\vmin^{\rm endo}$ obtained from it by $\delta \to -\delta$.
The same generalized Navarro--Frenk--White profile of Ref.~\cite{Cholis:2021rpp} sets the LZ rate and all $\gamma$-ray signals.

We treat the LZ event as signal and fit it with the extended likelihood
\begin{align}
\ln\mathcal{L} =&\,
 \ln {\rm Pois}\!\left(1\,\vert\,N_{\roi}(\vec{\theta})\right)
+\ln {\rm Pois}\!\left(0\,\vert\,N_{\rm SB}(\vec{\theta})\right) \nonumber \\
&+\ln p(248\keV\,\vert\,\vec{\theta}),
\label{eq:likelihood}
\end{align}
where ${\rm Pois}$ is the Poisson probability, $N_{\roi}$ and $N_{\rm SB}$ are the expected numbers of signal events, $p$ is the normalized, efficiency-weighted ROI spectrum convolved with a Gaussian of width $32.5\keV$, the statistical and systematic uncertainties of the event in quadrature, and $\vec{\theta} = \{\mchi,\delta,\mzp\}$, with $\mphi = \mzp$ and $g_X$ fixed by the relic density.
Profiling $\mzp$ on each branch, we draw contours in the $(\mchi,\delta)$ plane at $\Delta(-2\ln\mathcal{L}) = 2.30$ and $6.18$, the $1\sigma$ and $2\sigma$ levels for two parameters in the asymptotic limit of Wilks' theorem~\cite{Wilks:1938dza}.
With a single event, this limit need not hold, so these contours indicate the preferred regions rather than exact confidence regions.

For $q^2 \lesssim \mzp^2$, $\mchi$ and $\delta$ determine the spectral shape, while $\mzp$ controls the normalization through $\sigma_{\chi p}$ and $f_2$, giving rise to a degeneracy in $\mzp$.
For small $\mzp$, efficient conversion and fast decays, $\Gamma_{\chi_2} \propto \mzp^{-4}$, suppress $f_2$, so that $N_{\roi} \ll 1$ despite the large cross section.
As $\mzp$ increases, $N_{\roi}$ initially rises because the increase in $f_2$ outweighs the decrease in $\sigma_{\chi p}$.
Once $f_2$ approaches $1/2$, the continued decrease in $\sigma_{\chi p}$ causes $N_{\roi}$ to fall.
The spectral term primarily constrains $\er^*$, while the Poisson term favors $N_{\roi} \simeq 1$.
This event count can be reached at two values of $\mzp$, one on the rising side and the other on the falling side of the curve (\cref{sfig:ladder}).
These two solutions define the low- and high-$\mzp$ branches, respectively (\cref{fig:allowed}).
For $\epsilon=10^{-4}$, the LZ best fit lies at $\mchi = 43.9\GeV$ and $\delta = 1\MeV$ on both branches, at $(\mzp, f_2) = (1.05\GeV, 1.5\times10^{-6})$ and $(25.1\GeV, 0.497)$, with $N_{\roi} \simeq 0.94$ and $-2\ln\mathcal{L}$ differing by less than $0.01$ (Table~\ref{stab:benchmarks}).

Unlike endothermic scattering, the signal does not rely on the high-speed tail of the DM velocity distribution ~\cite{Das:2026jsq}.
Within $32.5\keV$ of the event, $\vmin \le 241~{\rm km/s}$, far below the Earth-frame maximum of about $800~{\rm km/s}$, and adding the Large Magellanic Cloud component~\cite{Smith-Orlik:2023kyl,Rodd:2026tyn} changes $-2\ln\mathcal{L}$ at both best fits by less than $0.02$.

Since $\sigma_{\chi p}$ and $\Gamma_{\chi_2}$ both scale as $g_X^2\epsilon^2/\mzp^4$, while $g_X$ and the conversion cross section do not depend on $\epsilon$, the preferred $\mzp$ scales approximately as $\epsilon^{1/2}$.
With the same fit quality, the low-branch best fit moves from $1.05$ to $3.12\GeV$ for $\epsilon = 10^{-3}$, and the branches lie at $0.37$ and $7.6\GeV$ for $\epsilon = 10^{-5}$.

So far we have set $\mphi = \mzp$. To test whether the LZ fit requires the dark Higgs boson, we raise $\mphi$ to $10\,\mchi$, which closes $\chi_1\chi_2 \to Z'\phi$.
The relic density then requires a larger $g_X$, and the high-branch LZ best fit shifts only in $\mzp$, from $25.1$ to $29.0\GeV$, with the same $-2\ln\mathcal{L}$.
Such a heavy $\phi$ violates perturbative unitarity~\cite{Kahlhoefer:2015bea}, so this case serves only as a proxy for the model without $\phi$.
Still, the LZ fit alone may not require the $\chi_1\chi_2 \to Z'\phi$ channel.

\dfsection{Galactic Center $\gamma$-ray excess}
We compare both branches with the GCE observed by the Fermi-LAT~\cite{Goodenough:2009gk,Daylan:2014rsa,Fermi-LAT:2015sau,Cholis:2021rpp}.
As recommended by its authors, we use the southern-sky spectrum of Ref.~\cite{Cholis:2021rpp} with its systematic covariance.
The full $40^\circ\times40^\circ$ window and other data sets are discussed in \cref{ssec:gce}.

On the low branch, $f_2 \sim 10^{-6}$, so only $\chi_1\chi_1 \to Z'Z'$ contributes.
The $Z'$ decays into leptons and light mesons, and the boosted photon spectrum peaks near $10\GeV$ (orange in \cref{fig:lz-best-fit}).
This spectrum gives $\chi^2/{\rm dof} = 160.8/14$ (Table~\ref{stab:gce}), worse than $\chi^2 = 130.7$ for no signal.
The GCE therefore disfavors the low branch.

On the high branch, $f_2 \simeq 1/2$, so $\chi_2\chi_2 \to Z'Z'$ and the coannihilation $\chi_1\chi_2 \to Z'\phi$ also contribute.
Both $Z'$ and $\phi$ decay mainly into hadrons, and $\phi$ decays mostly into $b\bar b$. Without any fit to the GCE, the prompt $\gamma$-ray flux of the high-branch LZ best fit (\cref{fig:lz-best-fit}, right) gives $\chi^2/{\rm dof} = 18.6/14$ ($p = 0.18$).
This is comparable to the $b\bar b$ fit of Ref.~\cite{Cholis:2021rpp}, which frees the DM mass and cross section, with $17.1/12$ ($p = 0.15$).
Fitting a free flux normalization $A$ gives $A = 1.06 \pm 0.10$, so the cross section fixed by the relic density already reproduces the observed flux.

In addition, the GCE selects the DM mass, which the LZ event leaves open.
The LZ likelihood alone has a valley of nearly constant $\er^*$ that extends to TeV DM masses (\cref{fig:allowed}).
On the high branch, $-2\ln\mathcal{L}$ rises by only $1.6$ up to $\mchi \simeq 1\TeV$.
Heavier DM, however, gives a lower $\gamma$-ray flux with a harder spectrum, and the GCE fit worsens to $\chi^2/{\rm dof} = 100/14$ already at $\mchi \simeq 100\GeV$.
In a joint fit of the LZ event and the GCE with $\mphi = \mzp$, the best fit therefore lies at the low-mass end of the valley, set by $\delta < 2m_e$, with $\mchi \approx 40$--$44\GeV$ at the nominal $1\sigma$ level.

Similarly, the GCE constrains $\mphi$.
Letting $\mphi$ vary as well, the joint fit gives $\mphi \approx 16$--$39\GeV$ at the nominal $1\sigma$ level (\cref{ssec:gce}).
A much lighter $\phi$ gives a photon spectrum too hard for the GCE, and a much heavier one closes $\chi_1\chi_2 \to Z'\phi$.
For $\mphi = 10\,\mchi$ (green curve in \cref{fig:lz-best-fit}), $\chi^2/{\rm dof}$ rises to $32.9/14$ ($p = 0.003$).
Freeing $A$ lowers it only to $31.1/13$, so the mismatch lies in the spectral shape rather than in the flux level.
Without $\phi$, the high branch can hardly explain the GCE, whereas it may still explain the LZ event.

\dfsection{Other constraints}
Details are given in \cref{ssec:constraints}.
The $3\gamma$ flux from $\chi_2 \to \chi_1 + 3\gamma$, constrained by INTEGRAL/SPI~\cite{Krnjaic:2025zjl}, shares the factor $g_X^2\epsilon^2/\mzp^4$ with the LZ rate and is therefore fixed by it for given $\mchi$ and $\delta$, up to the $Z'$ propagator~\cite{Deka:2026zeu}.
Evaluated for our halo, the limit excludes the LZ best fits of \cref{fig:lz-best-fit} by factors of $1.5$ (high branch) and $1.7$ (low branch).
Since the decay width grows steeply with $\delta$, a slightly smaller $\delta = 0.99\MeV$ satisfies the limit with $\Delta(-2\ln\mathcal{L}) = 0.06$ and $0.09$, respectively (Table~\ref{stab:benchmarks}), so we do not impose it in \cref{fig:allowed,fig:lz-best-fit}.
There the high-branch GCE fit gives $\chi^2/{\rm dof} = 18.1/14$.

The same decays inject energy after recombination, constrained by the CMB anisotropies~\cite{Slatyer:2016qyl}.
On the low branch, $f_2$ is the remnant of a much larger pre-decay fraction $f_2^{\rm conv}$ (Table~\ref{stab:benchmarks}). At the best fit, the energy injected during the dark ages exceeds the limit by a factor of $53$, and still by about $13$ for the least sensitive injection spectrum. The limit excludes $73\%$ of the low-branch region inside the $1\sigma$ contour of \cref{fig:allowed}. We do not assess the rest, at heavier DM and lighter $Z'$, where annihilation and big bang nucleosynthesis (BBN) limits would also apply.
On the high branch, the injection lies more than two orders of magnitude below the limit.

High-energy solar neutrino searches~\cite{Super-Kamiokande:2015xms,ANTARES:2016xuh,IceCube:2025fcu} exclude neither LZ best fit.
On the low branch, the muons and mesons from $Z'$ decays stop in the Sun, so that their neutrinos fall below the search thresholds.
At the high-branch best fit, the solar capture rate corresponds to an effective spin-dependent cross section $\sigma^{\rm eff}_\odot = 5.0\times10^{-44}~\mathrm{cm}^2$, about 600 times below the IceCube $\tau^+\tau^-$ limit.
Only for $\mchi \gtrsim 400\GeV$ does part of the high-branch region reach these limits (\cref{ssec:constraints}).

The mediators $Z'$ and $\phi$ are also constrained by collider searches and by BBN.
Since $Z' \to \chi_1\chi_2$ is kinematically closed, the $Z'$ decays into SM particles, as a minimal dark photon does. Its coupling, $\epsilon c_W e \simeq 2.7\times10^{-5}$, lies about 8 times below the BaBar limit~\cite{BaBar:2014zli} and outside the preliminary FASER limit~\cite{faser_collaboration_2026_fwshk-fak94} near the low-branch best fit.
At the high-branch best fit, $25.1\GeV$, it lies 23--25 times below the LHCb and CMS dimuon limits~\cite{LHCb:2019vmc,CMS:2019buh}.
With $\sa = 10^{-6}$, $\phi$ decays before BBN at our best fits~\cite{Winkler:2018qyg,Ferber:2023iso}.

\dfsection{Conclusion}
We have computed $f_2$ from the thermal history rather than fixing it.
Since $\mzp$ then sets both $f_2$ and the scattering rate, two branches of $\mzp$ can explain the LZ event for a given $\epsilon$. CMB limits and the GCE disfavor the low branch.

On the high branch, the southern-sky GCE complements the LZ event.
It selects $\mchi \approx 40$--$44\GeV$, which the LZ event alone leaves open, and the relic cross section reproduces the observed flux.
The dark Higgs boson thus makes the model consistent and, with a mass near that of the dark photon, also helps reproduce the GCE spectrum.

Additional exposure, from further LZ data taking or from existing XENONnT and PandaX-4T data in an extended recoil window, can test both the LZ event itself and our interpretation of it~\cite{Dent:2026bji}.
Future argon detectors would give a cleaner test, since the signal there lies below the first zero of the form factor~\cite{deLima:2026shq,Baer:2026fpy,Baer:2026yrt}.

\begin{acknowledgments}
This work is supported by the National Research Foundation of Korea (NRF) Research Grants No. RS-2024-00341419 (J.K.K.) and No. RS-2025-24803289 (P.K.). The work is also supported by KIAS Individual Grants No. PG099202 (J.H.K.) and No. QP021402 (P.K.).
\end{acknowledgments}

\putbib[LZ2026,LZ2026-aux]
\makeatletter
\global\let\@FMN@list\@empty
\global\let\auto@bib\@empty
\makeatother
\end{bibunit}


\begin{bibunit}[apsrev4-2]
\clearpage
\onecolumngrid
\setcounter{equation}{0}
\setcounter{figure}{0}
\setcounter{table}{0}
\setcounter{secnumdepth}{1}
\setcounter{section}{0}
\renewcommand{\thesection}{S\arabic{section}}
\setcounter{page}{1}
\makeatletter
\renewcommand{\theequation}{S\arabic{equation}}
\renewcommand{\thefigure}{S\arabic{figure}}
\renewcommand{\thetable}{S\arabic{table}}
\renewcommand{\theHequation}{S\arabic{equation}}
\renewcommand{\theHfigure}{S\arabic{figure}}
\renewcommand{\theHtable}{S\arabic{table}}
\renewcommand{\bibnumfmt}[1]{[S#1]}
\renewcommand{\citenumfont}[1]{S#1}

\let\oldthebibliography\thebibliography
\renewcommand{\thebibliography}[1]{%
  \oldthebibliography{#1}%
  \setcounter{NAT@ctr}{0}
  \setcounter{enumiv}{0}
}
\makeatother

\begin{center}
  \textbf{\large Supplemental Material: Exothermic Dark Matter with a Dark Higgs:\\ The LZ 248 keV Event and the Galactic Center Excess}\\[0.5em]
  \text{Jinheung\,\,Kim, Jongkuk\,\,Kim, and Pyungwon\,\,Ko}
\end{center}
\vspace{1em}

\section{Model}\label{ssec:model}

\begin{figure}[!hb]
\centering
\begin{minipage}{0.32\linewidth}\centering
\includegraphics[width=\linewidth]{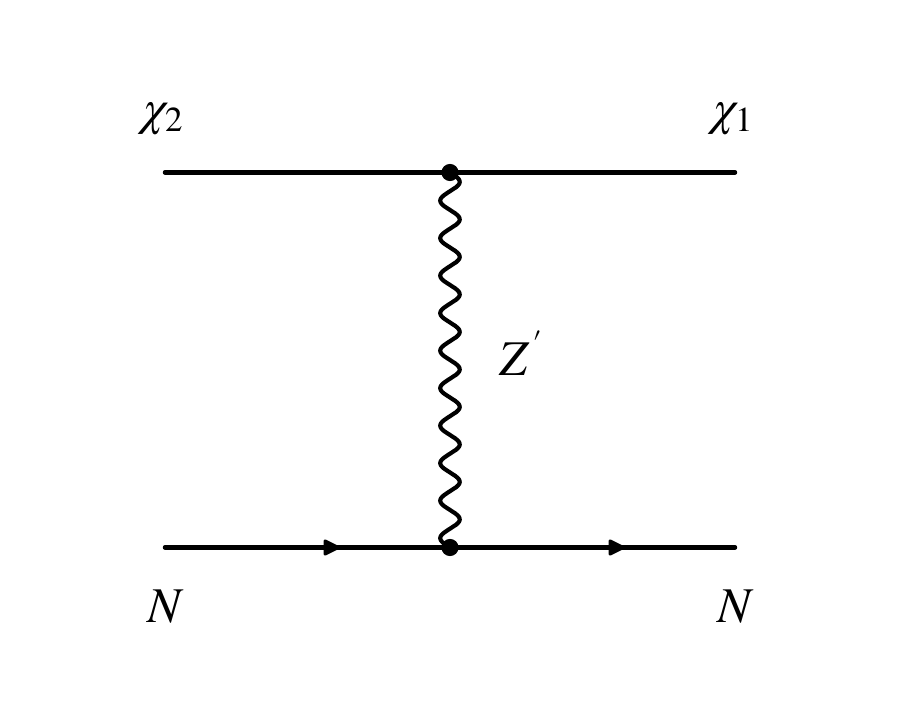}\\[-2pt](a)
\end{minipage}\hfill
\begin{minipage}{0.32\linewidth}\centering
\includegraphics[width=\linewidth]{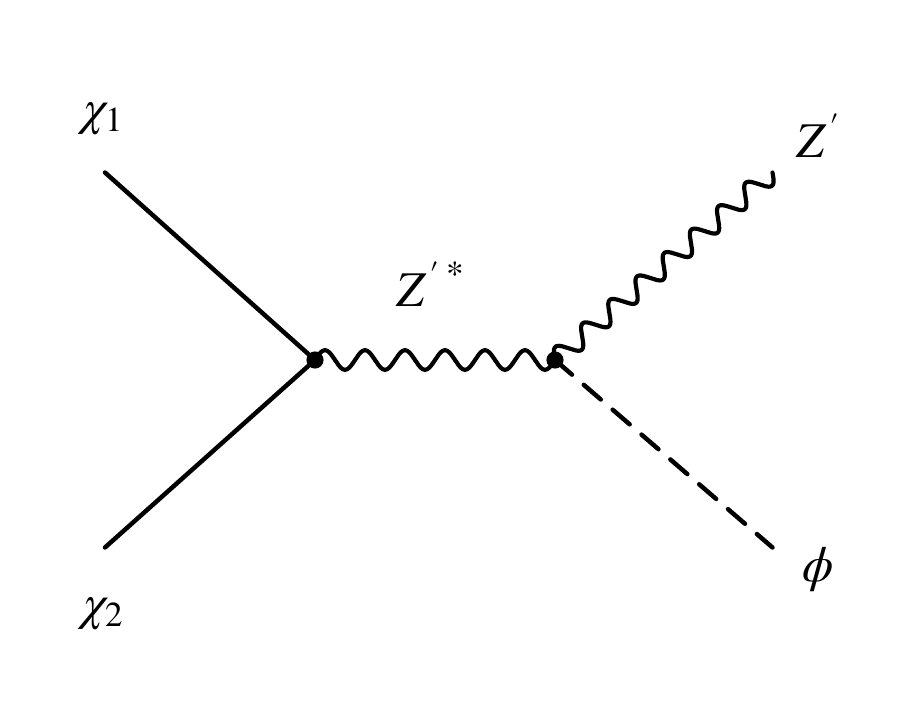}\\[-2pt](b)
\end{minipage}\hfill
\begin{minipage}{0.32\linewidth}\centering
\includegraphics[width=\linewidth]{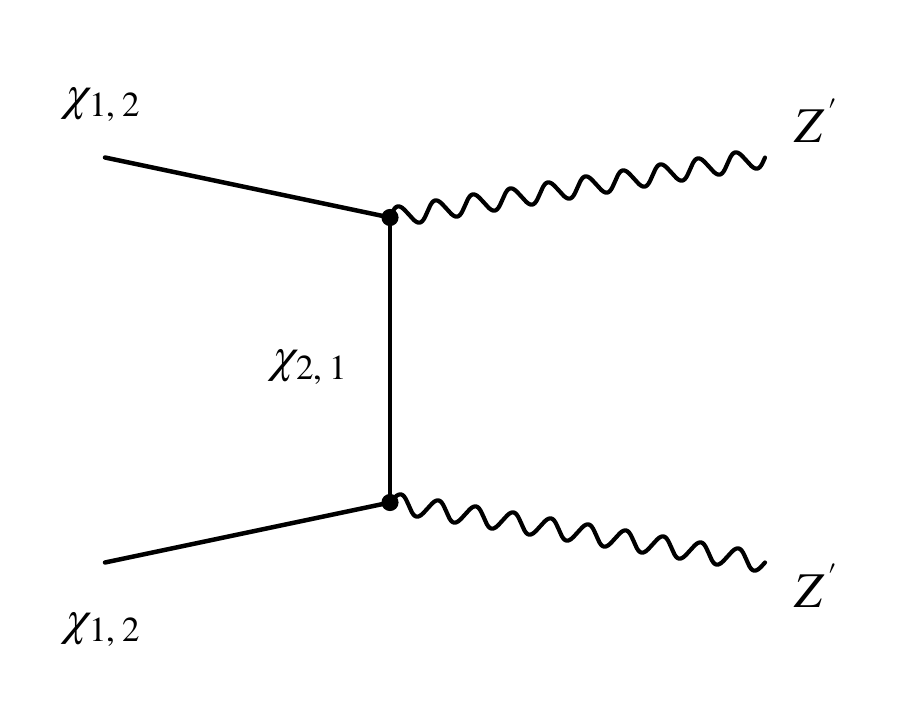}\\[-2pt](c)
\end{minipage}
\caption{Leading diagrams of the processes in our analysis.
(a) Exothermic scattering $\chi_2 N \to \chi_1 N$ through $t$-channel $Z'$ exchange, with $N$ the target nucleus.
(b) Coannihilation $\chi_1\chi_2 \to Z'\phi$ through $s$-channel $Z'$ exchange (dark Higgsstrahlung).
(c) Annihilation $\chi_i\chi_i \to Z'Z'$ through $t$-channel exchange of the other Majorana state, $\chi_2$ for $i = 1$ and $\chi_1$ for $i = 2$.
The $u$-channel diagram and the $s$-channel $\phi$ exchange of this process, discussed below, are not shown.
Wavy lines denote the $Z'$ and the dashed line the $\phi$.}
\label{sfig:diagrams}
\end{figure}

Our model~\cite{Baek:2020owl} has a dark $U(1)_X$ vector boson, a Dirac fermion $\chi$ with dark charge $+1$, and a dark Higgs field $\Phi$ with dark charge $+2$, described by the renormalizable Lagrangian
\begin{align}
\mathcal{L}
=&\,
\mathcal{L}_{\rm SM}
-\frac{1}{4}X_{\mu\nu}X^{\mu\nu}
-\frac{\sin\epsilon}{2}
 X_{\mu\nu}B^{\mu\nu} +\bar{\chi}(i\slashed{D} - m_D)\chi
-\frac{1}{\sqrt{2}}\left(y\,\Phi^\dagger \overline{\chi^c}\chi + \mathrm{H.c.}\right)
\nonumber\\
&+(D_\mu\Phi)^\dagger(D^\mu\Phi)
+m_\Phi^2\Phi^\dagger\Phi
-\lambda_\Phi(\Phi^\dagger\Phi)^2
-\lambda_{H\Phi}(H^\dagger H)(\Phi^\dagger\Phi).
\label{seq:lagrangian}
\end{align}
The dark Higgs field develops a vacuum expectation value, $v_\Phi \approx (m_\Phi^2/\lambda_\Phi)^{1/2}$ when $\lambda_{H\Phi}$ is small, and breaks $U(1)_X$.
In the unitary gauge, $\Phi = (v_\Phi + \phi_0)/\sqrt{2}$, which gives the $X$-boson mass $m_X = 2g_Xv_\Phi$, the mass splitting $\delta = 2yv_\Phi$, and $\mphi^2 \simeq 2\lambda_\Phi v_\Phi^2$, so that $\lambda_\Phi \simeq 2g_X^2\mphi^2/\mzp^2$.
The breaking of $U(1)_X$ leaves a local $Z_2$ symmetry under 
which only $\chi$ is odd, so that $\chi_1$ is stable.
The portal coupling $\lambda_{H\Phi}$ mixes $\phi_0$ with the SM Higgs boson into the mass eigenstates $\phi$ and $h$, the observed Higgs boson, with mixing angle $\alpha$ ($\sa \equiv \sin\alpha$).
\Cref{sfig:diagrams} shows the leading diagrams of the processes that set the LZ signal and the relic density.

The kinetic mixing shifts the $Z'$ mass only at $\calO(\epsilon^2)$, $\mzp^2 = m_X^2\left[1+\calO(\epsilon^2)\right]$.
For $\mzp \ll m_Z$, it also suppresses the weak-neutral-current coupling of the $Z'$ by $(\mzp/m_Z)^2$ relative to the electromagnetic coupling $\epsilon c_W e$.
In direct detection, however, the momentum transfer is small.
The exchanges of both neutral mass eigenstates then add up to $\epsilon c_W e Q_f/\mzp^2$ for any $\mzp < m_Z$, so that the DM--nucleus cross section keeps this form up to $\mzp$ close to $m_Z$.
We nevertheless compute the $\chi_2$ decay width and the kinetic-decoupling rates with the exact diagonalization of the neutral gauge-boson mass matrix.
For these, the $Z$ mixing cannot be neglected, since it gives the couplings to neutrinos and the axial couplings to leptons.

Previous exothermic explanations of the LZ event with a dark photon~\cite{deLima:2026shq,Baer:2026fpy,Deka:2026zeu} use a model without a dark Higgs field.
There, the $Z'$ mass can come from the St\"{u}ckelberg mechanism or a Proca term, and the splitting from a dimension-three soft-breaking term, as in Ref.~\cite{Baer:2026fpy}:
\begin{equation}
\calL ({\rm no}~ \Phi ) \supset \frac{1}{2} m_{Z'}^2 Z'_\mu Z'^\mu -
\frac{1}{4} \delta  ( \overline{\chi^c}\chi + \mathrm{H.c.}) .
\label{seq:simplified}
\end{equation}
We call this the simplified (Higgsless) model.
It describes the exothermic LZ kinematics equally well, since these depend only on $\mchi$ and $\delta$, but it differs from our model in two respects.
Phenomenologically, it lacks the channels with $\phi$.
In the relic density, $\chi_1\chi_2 \to Z'\phi$ contributes comparably to $\chi_i\chi_i \to Z'Z'$ for $\mphi = \mzp$, whereas $\chi_i\chi_i \to \phi\phi$ and $\chi_i\chi_i \to \phi^* \to f\bar f$ are suppressed by $\delta/\mzp$, the latter also by $\sa$.
We approximate the simplified model by the decoupling limit of $\phi$, with $\mphi = 10\,\mchi$ as a proxy.
Since the annihilation cross sections scale as $g_X^4$, closing $\chi_1\chi_2 \to Z'\phi$ there raises the LZ best-fit coupling from $g_X = 0.111$ to $0.149$, a factor of $1.34$ at a slightly heavier $Z'$ (Table~\ref{stab:benchmarks}).
The same channel also plays a key role for the GCE, since $\phi \to b\bar b$ changes the shape of the $\gamma$-ray spectrum, and without it the GCE fit worsens even with a free normalization, as shown in the main text.

Theoretically, the dimension-three term breaks $U(1)_X$ explicitly.
The simplified model is therefore not gauge invariant, which raises two issues.

First, no symmetry then forbids other gauge-noninvariant terms.
An example is $\lambda_{Z'H} Z'_\mu Z'^\mu H^\dagger H$, which induces $h \to Z'Z'$ with the width~\cite{Baek:2021hnl}
\begin{equation}
\Gamma(h \to Z'Z') = \frac{\lambda_{Z'H}^2 v_H^2 m_h^3}{32\pi\,\mzp^4}
\left(1-\frac{4\mzp^2}{m_h^2}+\frac{12\mzp^4}{m_h^4}\right)
\left(1-\frac{4\mzp^2}{m_h^2}\right)^{1/2}.
\label{seq:hzpzp}
\end{equation}
Here $v_H$ is the Higgs vacuum expectation value.
For fixed $\lambda_{Z'H}$, this width diverges as $\mzp \to 0$ through the longitudinal $Z'$ bosons, whose growth no gauge symmetry cancels.
This infrared breakdown makes the simplified model a poor effective theory.
Avoiding it requires setting $\lambda_{Z'H}$ to zero by hand, with no symmetry reason to do so.

In our model, however, gauge invariance forbids a direct $Z'_\mu Z'^\mu H^\dagger H$ term.
The $hZ'Z'$ coupling arises only through the $h$--$\phi_0$ mixing and is tied to the $Z'$ mass, as in the renormalizable Higgs-portal vector DM model~\cite{Farzan:2012hh, Baek:2012se}.
The resulting coupling, $|\lambda_{Z'H}| = |\sa|\,\mzp^2/(v_H v_\Phi)$ in \cref{seq:hzpzp}, cancels the $1/\mzp^4$ divergence.
For $\mzp \ll m_h$, the longitudinal $Z'$ bosons act as the Goldstone mode of $\Phi$, by the equivalence theorem.
The width then approaches the finite value $\Gamma(h \to Z'Z') \simeq \sa^2 m_h^3/(32\pi v_\Phi^2)$ of $h$ decaying into this mode~\cite{Baek:2021hnl}.
This $\mzp^2$ scaling of $\lambda_{Z'H}$ is also what Higgs-width measurements~\cite{ATLAS:2018jym,CMS:2019ekd} require for $\mzp \ll m_h$.
For our $\sa = 10^{-6}$, the factor $\sa^2 = 10^{-12}$ makes this width negligible.

\begin{figure}[t]
    \centering
    \includegraphics[width=0.49\linewidth]{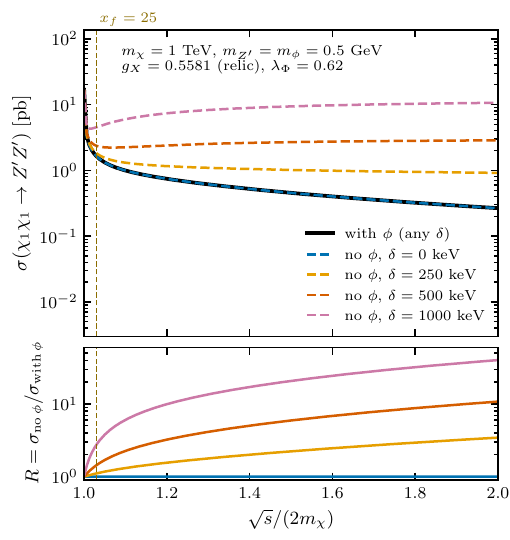}
    \includegraphics[width=0.49\linewidth]{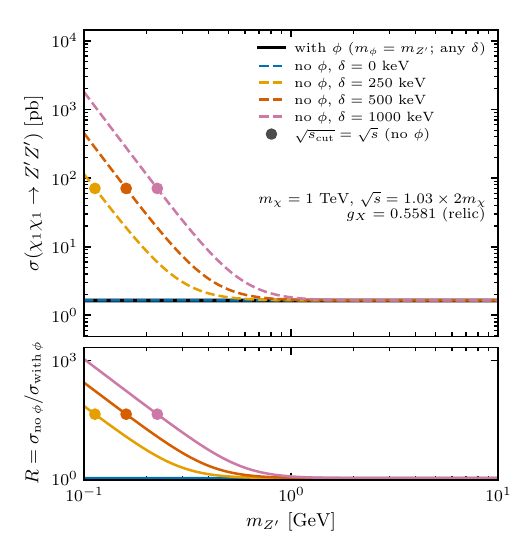}
    \caption{Left: Cross section for $\chi_1\chi_1 \to \zp\zp$ as a function of $\sqrt{s}/(2\mchi)$ in our model (with $\phi$; FDM denotes the fermion DM model) and in the simplified model (no $\phi$), for several values of $\delta$, with $\mchi = 1\TeV$ and $\mzp = \mphi = 0.5\GeV$. The vertical line marks $\sqrt{s}/(2\mchi) = 1.03$, typical of freeze-out ($x_f = 25$). Right: The same cross section as a function of $\mzp$ for $\sqrt{s}/(2\mchi) = 1.03$. The circles mark the points where $\sqrt{s} = \sqrt{s_{\rm cut}}$.}
    \label{sfig:unitarity}
\end{figure}%

Second, in the simplified model, the amplitudes of $\chi_i Z' \to \chi_i Z'$ and $\chi_i\chi_i \to Z'Z'$ with longitudinal $Z'$ bosons grow with energy for $\delta \neq 0$ and eventually violate unitarity~\cite{Ko:2019wxq,Jeong:2026dzt}.
We take $\chi_1\chi_1 \to Z'Z'$ as an example, for which
\begin{equation}
|\calM(\chi_1\chi_1\to \zp_L \zp_L)| \sim \frac{2g_X^2\,\delta}{\mzp^2}\sqrt{s}.
\label{seq:amplitude}
\end{equation}
In our model, the $\phi$ exchange cancels this growth, in the $s$ channel for $\chi_i\chi_i \to Z'Z'$ and in the $t$ channel for $\chi_i Z' \to \chi_i Z'$.
The $J = 0$ partial-wave bound $|{\rm Re}\,a_0| < 1/2$ gives the high-energy estimate $\sqrt{s_{\rm cut}} \simeq 4\pi\mzp^2/(g_X^2\delta)$.
Above $\sqrt{s_{\rm cut}}$, the simplified model loses perturbative consistency.
Already at $\sqrt{s} \sim \mzp^2/\delta$, a factor of $4\pi/g_X^2$ below $\sqrt{s_{\rm cut}}$, the energy-growing term becomes comparable to the energy-independent $\calO(g_X^2)$ part of the amplitude.
Above this energy, the two models can differ significantly.

\Cref{sfig:unitarity} (left) shows this. In our model, the cross section falls with $\sqrt{s}$. In the simplified model, it also falls for $\delta = 0$, but for $\delta \neq 0$ it levels off at a value that grows with $\delta$.
At $\sqrt{s} = 1.03 \times 2\mchi$, typical of freeze-out, the simplified model gives a cross section larger than ours by up to a factor of about 40 below $\sqrt{s_{\rm cut}}$, growing as $\mzp$ decreases (\cref{sfig:unitarity}, right).
It therefore also gives a different relic density.

\Cref{sfig:unitarity} uses $\mchi = 1\TeV$ and $g_X = 0.56$.
For $\mzp = 0.5\GeV$ and $\delta = 1\MeV$ (left panel), $2\mchi\delta/\mzp^2 = 8$, so that the energy-growing term already dominates the amplitude near threshold.
At our best fits, in contrast, $2\mchi\delta/\mzp^2$ is only $0.08$ (low branch) and $1.4\times10^{-4}$ (high branch), and it stays below $0.04$ over the whole high-branch region of \cref{fig:allowed}.
On the low branch, however, the $Z'$ becomes lighter toward heavier DM and smaller $\delta$.
In this direction, $2\mchi\delta/\mzp^2$ exceeds one even inside the $1\sigma$ region, wherever $\delta \lesssim 700\keV$.
This part of the region begins at $\mchi \simeq 50\GeV$, just above the best fit.
There, the two models can differ significantly even in the relic density.

Both issues arise from the explicit breaking of $U(1)_X$, and both are resolved by introducing the dark Higgs field $\Phi$, as in our model.
In principle, the simplified model is recovered in the decoupling limit $\mphi \to \infty$ of our model.
Since $\lambda_\Phi \simeq 2g_X^2\mphi^2/\mzp^2$, however, perturbative unitarity bounds $\mphi \lesssim \sqrt{\pi}\,\mzp/g_X$~\cite{Kahlhoefer:2015bea}, so this limit cannot be reached.
A consistent description therefore requires the dark Higgs boson.

Inelastic scalar DM in a similar dark $U(1)_X$ model could also explain the LZ event, since the exothermic kinematics depends only on $\mchi$ and $\delta$, but with a different excited-state fraction, relic density, and several additional quartic couplings. Such models have been studied by one of us for the GCE~\cite{Baek:2014kna} and by two of us in $U(1)_{L_\mu - L_\tau}$~\cite{Baek:2022ozm}. We leave this case for future work.

\section{Thermal history of $\chi_1$ and $\chi_2$}\label{ssec:history}
The relic density is set by $\chi_i\chi_i \to Z'Z'$ ($i = 1,2$) and $\chi_1\chi_2 \to Z'\phi$, followed by the $Z'$ and $\phi$ decays through the kinetic and Higgs mixing. The channels $\chi_i\chi_i \to \phi\phi,\, hh$ and $\chi_1\chi_2 \to Z'^{(*)} \to f\bar{f}$, also included, are negligible owing to the suppression by $\delta/\mzp$, $\sa$, and $\epsilon$, respectively.
For the LZ best fits with $\epsilon=10^{-4}$ and $\sa=10^{-6}$, the $Z'$ and $\phi$ decay before the onset of big bang nucleosynthesis (BBN)~\cite{Winkler:2018qyg,Ferber:2023iso}. A lighter $\phi$, below the dimuon threshold, would be long-lived for this $\sa$ and thus constrained by BBN.

\begin{table}[t]
\centering
\caption{LZ-only best fits and INTEGRAL-allowed points for $\epsilon = 10^{-4}$, $\sa = 10^{-6}$, and $\rho_\odot = 0.4\GeV/{\rm cm}^3$.
The first three columns are the best fits to the LZ data alone (\cref{fig:allowed,fig:lz-best-fit}), and the last two are the best fits allowed by INTEGRAL at $(\mchi,\delta) = (43.9\GeV, 0.99\MeV)$.
$\Phi_{3\gamma}/\Phi_{3\gamma}^{\rm lim}$ is the $3\gamma$ flux in units of the INTEGRAL limit.
$A$ is the fitted normalization of the $\gamma$-ray flux, which is not restricted to positive values. All other quantities are defined in the text.}
\label{stab:benchmarks}
\begin{tabular}{lccccc}
\toprule
 & \multicolumn{3}{c}{LZ only} & \multicolumn{2}{c}{INTEGRAL allowed} \\
 & low & high & high, $\mphi = 10\,\mchi$ & low & high \\
\midrule
$\mchi$ [GeV] & 43.9 & 43.9 & 43.9 & 43.9 & 43.9 \\
$\delta$ [keV] & 1000 & 1000 & 1000 & 990 & 990 \\
$\mzp$ [GeV] & 1.050 & 25.08 & 29.00 & 0.986 & 26.33 \\
$\mphi$ [GeV] & 1.050 & 25.08 & 439.1 & 0.986 & 26.33 \\
$g_X$ & 0.1162 & 0.1112 & 0.1488 & 0.1162 & 0.1109 \\
$\lambda_\Phi$ & 0.0270 & 0.0247 & 10.2 & 0.0270 & 0.0246 \\
$\tkd$ [GeV] & 0.0242 & 0.618 & 0.614 & 0.0229 & 0.654 \\
$f_2^{\rm conv}$ & 0.0285 & 0.497 & 0.496 & 0.0252 & 0.497 \\
$f_2$ & $1.54\times10^{-6}$ & 0.4971 & 0.4957 & $8.89\times10^{-7}$ & 0.4974 \\
$\tau_{\chi_2}$ [s] & $4.43\times10^{16}$ & $1.57\times10^{22}$ & $1.57\times10^{22}$ & $4.24\times10^{16}$ & $2.37\times10^{22}$ \\
$\tau_{\chi_2}\mchi$ [s GeV] & $1.94\times10^{18}$ & $6.90\times10^{23}$ & $6.89\times10^{23}$ & $1.86\times10^{18}$ & $1.04\times10^{24}$ \\
$\Phi_{3\gamma}/\Phi_{3\gamma}^{\rm lim}$ & 1.67 & 1.52 & 1.52 & 1.00 & 1.00 \\
$N_{\roi}$ & 0.944 & 0.941 & 0.941 & 0.757 & 0.838 \\
$N_{\rm SB}$ & 0.0550 & 0.0587 & 0.0587 & 0.0367 & 0.0439 \\
$-2\ln\mathcal{L}$ & 12.342 & 12.339 & 12.339 & 12.436 & 12.396 \\
$\langle\sigma v\rangle_{\rm eff}$ [$10^{-26}~{\rm cm^3/s}$] & 2.20 & 2.32 & 2.25 & 2.20 & 2.30 \\
$\chi^2_{\rm GCE}/{\rm dof}$, south & 160.8/14 & 18.6/14 & 32.9/14 & 153.6/14 & 18.1/14 \\
$\chi^2_{\rm GCE}/{\rm dof}$, full & 250.6/14 & 51.8/14 & 67.9/14 & 225.1/14 & 51.6/14 \\
$\chi^2_{\rm GCE}/{\rm dof}$, Di Mauro~\cite{DiMauro:2026fnp} & 257.6/11 & 44.1/11 & 69.3/11 & 432.1/11 & 41.1/11 \\
$A$, south & $0.16 \pm 0.15$ & $1.06 \pm 0.10$ & $1.16 \pm 0.12$ & $-0.06 \pm 0.22$ & $1.07 \pm 0.10$ \\
$A$, full & $-0.23 \pm 0.13$ & $0.98 \pm 0.09$ & $1.08 \pm 0.11$ & $-0.56 \pm 0.19$ & $0.98 \pm 0.09$ \\
$\sigma^{\rm eff}_\odot$ [$10^{-44}~{\rm cm^2}$] & 5.43 & 5.02 & --- & 4.04 & 4.12 \\
CMB ratio & 53.0 & 0.0023 & 0.0023 & 48.8 & 0.0015 \\
\bottomrule
\end{tabular}
\end{table}
\begin{figure}[!htb]
    \centering
    \includegraphics[width=0.62\linewidth]{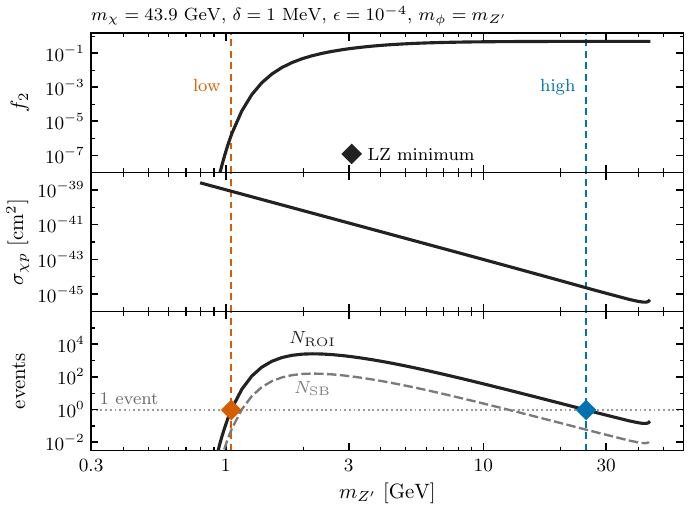}
    \caption{The excited-state fraction $f_2$ (top), the DM--proton cross section $\sigma_{\chi p}$ (middle), and the expected numbers of signal events in the ROI and the SB (bottom) as functions of $\mzp$ at $(\mchi,\delta) = (43.9\GeV, 1\MeV)$, for $\epsilon = 10^{-4}$, $\mphi = \mzp$, $g_X$ fixed by the relic density, and $\rho_\odot = 0.4\GeV/{\rm cm}^3$.
    The dotted line marks one event, and diamonds and dashed lines mark the LZ best fits of the low (orange) and high (blue) branches.}
    \label{sfig:ladder}
\end{figure}

After chemical freeze-out, the total DM number density $n_\chi = n_{\chi_1}+n_{\chi_2}$ only dilutes as $a^{-3}$, and the fractions $f_i = n_{\chi_i}/n_\chi$ evolve through the conversion $\chi_2\chi_2 \leftrightarrow \chi_1\chi_1$~\cite{Batell:2009vb,Baryakhtar:2020rwy,CarrilloGonzalez:2021lxm}, the scattering $\chi_2\ell \leftrightarrow \chi_1\ell$ on the bath, and the decays $\chi_2 \to \chi_1 + \SM$.

Since $\delta$ is well below the freeze-out temperature $T_{\rm fo}$, $f_1 \simeq f_2 \simeq 1/2$ at chemical freeze-out.
The conversion can then reduce $f_2$ once $\tchi$ drops below $\delta$, especially for small $\mzp$. The fraction $f_2$ evolves as
\begin{align}
\frac{df_2}{dt} =&\, -n_\chi\expval{\sigma_{22\to11}v}\left[f_2^2 - \xi^2(\tchi)\,(1-f_2)^2\right]
\nonumber\\
&- \Gamma_\downarrow(\tsm)\, f_2 + \Gamma_\uparrow(\tsm)\,(1-f_2) - \Gamma_{\chi_2} f_2,
\label{seq:f2-boltzmann}
\end{align}
where  $\xi(T) \equiv (1+\delta/\mchi)^{3/2}e^{-\delta/T}$ is the equilibrium ratio $n_{\chi_2}/n_{\chi_1}$ at temperature $T$.
The first term describes conversion at the DM temperature $\tchi$, with the inverse rate fixed by detailed balance.
The second and third terms describe scattering $\chi_2\ell \leftrightarrow \chi_1\ell$ with charged leptons $\ell = e,\mu,\tau$. We compute the down-scattering rate $\Gamma_\downarrow$ from the same $Z'$-mediated amplitudes used for the momentum-transfer rate below.
By detailed balance, the up-scattering rate is $\Gamma_\uparrow = \xi(\tsm)\,\Gamma_\downarrow$, with $\xi$ evaluated at the bath temperature $\tsm$. The last term describes $\chi_2$ decays.
Since $\Gamma_{\chi_2}$ is negligible compared with the Hubble rate while the conversion and the bath scattering are active, we solve \cref{seq:f2-boltzmann} without the last term from chemical equilibrium, which gives the fraction $f_2^{\rm conv}$ after these processes freeze out. The present fraction is then $f_2 = f_2^{\rm conv}\exp(-t_0/\tau_{\chi_2})$, with $t_0$ the age of the Universe.

The DM temperature, $\tchi \equiv \frac{2}{3}\expval{\mathbf{p}^2/2\mchi}$, follows from the scattering $\chi_{1,2}\,\ell \to \chi_{2,1}\,\ell$, which keeps $\tchi = \tsm$ early on~\cite{Bringmann:2006mu,Bringmann:2009vf,Visinelli:2015eka}:
\begin{equation}
\frac{d\tchi}{dt} + 2H\tchi = 2\gamma(\tsm)\,(\tsm - \tchi),
\label{seq:tchi-evolution}
\end{equation}
with 
\begin{equation}
\gamma(T)=\sum_\ell \frac{g_\ell}{6\mchi T}\int\!\frac{d^3p}{(2\pi)^3}\,f_\ell(1-f_\ell)\,\frac{p}{E}\,\sigma_T(E),
\end{equation}
where $f_\ell$ is the Fermi--Dirac distribution of a lepton $\ell$ with momentum $p$ and energy $E$, $g_\ell = 4$ counts its spin and charge states, and $\sigma_T = \int d\sigma\,(-t)$ is the cross section of $\chi_{1,2}\ell\to\chi_{2,1}\ell$ weighted by the momentum transfer $-t$ and by the equilibrium populations of $\chi_1$ and $\chi_2$.
This $\gamma$ is half the rate of the same name in Ref.~\cite{Bringmann:2006mu}.
After the lepton scattering freezes out, DM without conversions would simply cool as
\begin{equation}
\tchi^{(0)}(\tsm) = \frac{\tsm^2}{\tkd} \left(\frac{g_{*s}(\tsm)}{g_{*s}(\tkd)}\right)^{2/3} \qquad (\tsm < \tkd),
\label{eq:tchi}
\end{equation}
which defines the kinetic decoupling temperature $\tkd$, here computed from the scattering on charged leptons alone.
Quarks and hadrons can be neglected, since the low-branch $\tkd$ lies well below the QCD crossover and the high-branch $f_2 \simeq 1/2$ is insensitive to $\tkd$.

The conversions, however, release $2\delta$ per $\chi_2\chi_2 \to \chi_1\chi_1$ and heat the dark sector after kinetic decoupling. We include this heating through entropy conservation in the dark sector, assuming a common $\chi_{1,2}$ temperature and conversions fast enough to keep $f_2 = f_2^{\rm eq}(\tchi)$ until they freeze out~\cite{DallaValleGarcia:2025cwf}: 
\begin{equation}
\tchi = \tchi^{(0)}\left(\frac{1-f_2^{\rm eq}}{1-f_2^{\rm kd}}\right)^{2/3}\exp\!\left[-\frac{2}{3}\left(\frac{f_2^{\rm eq}\,\delta}{\tchi}-\frac{f_2^{\rm kd}\,\delta}{\tkd}\right)\right],
\label{seq:tchi-reheated}
\end{equation}
where  $f_2^{\rm eq} = \xi(\tchi)/[1+\xi(\tchi)]$, and $f_2^{\rm kd}$ is its value at $\tkd$.
The equilibrium assumption suffices, since $\tchi$ no longer affects $f_2$ after the conversions freeze out.
The heating raises $\tchi$ above $\tchi^{(0)}$ by at most a factor of $2^{2/3}$, reached for $\delta \ll \tkd$ once most $\chi_2$ have converted, as on the low branch.
On the high branch, $f_2^{\rm conv} \simeq 0.497$ (Table~\ref{stab:benchmarks}) differs from $1/2$ by less than $1\%$, which bounds the effect of the conversions, and thus of this heating, on $f_2$.

For $\alpha_X\mchi/\mzp \gtrsim 1$, multiple $Z'$ exchanges can suppress the thermally averaged conversion rate relative to the Born approximation and make it oscillate with $\mzp$~\cite{Baer:2026fpy}. This regime occurs for heavy DM and a light $Z'$ on the low branch, whereas $\alpha_X\mchi/\mzp < 0.05$ at the LZ best fits.
We therefore compute $\sigma_{22\to11}$ by solving the coupled-channel Schr\"odinger equation in the basis $(\chi_1\chi_1,\,\chi_2\chi_2)$ with the potential~\cite{Ko:2019wxq,Baer:2026fpy}
\begin{equation}
\mathbf V(r)=
\begin{pmatrix}
0&-\alpha_X e^{-\mzp r}/r\\
-\alpha_X e^{-\mzp r}/r&2\delta
\end{pmatrix},
\label{eq:V-sommerfeld}
\end{equation}
with $\alpha_X = g_X^2/4\pi$, neglecting the $\phi$ contribution, suppressed by $\delta/\mzp$.
Partial wave by partial wave, we integrate the radial equations outward from the regular solution at the origin and impose an incoming wave only in the $\chi_2\chi_2$ channel, as in Refs.~\cite{Ko:2019wxq,Baer:2026fpy}. The coefficient $S^{(L)}_{12}$ of the flux-normalized outgoing $\chi_1\chi_1$ wave in partial wave $L$, with channel momenta $k_2 = \mu v$ and $k_1 = \sqrt{k_2^2 + 4\mu\delta}$, $\mu \simeq \mchi/2$, gives $\sigma_{22\to11}(v) = (\pi/k_2^2)\sum_L w_L\,(2L+1)\,|S^{(L)}_{12}|^2$, where $w_L = 1/2$ ($3/2$) for even (odd) $L$ accounts for the spin-singlet (triplet) states of the identical Majorana pairs.
We sum the partial waves until convergence and average $\sigma_{22\to11}v$ over a Maxwell distribution of relative velocities at $\tchi$.

After the conversion and the bath scattering freeze out, $f_2$ changes only through the $\chi_2$ decays, which conserve the total DM number owing to the residual $Z_2$ symmetry.
We restrict ourselves to $\delta \le 1\MeV$, below $2m_e$, since $\chi_2 \to \chi_1 e^+e^-$ would otherwise remove $\chi_2$ well before today.
Then only two decays remain, $\chi_2 \to \chi_1\nu\bar\nu$ through the $Z$--$Z'$ mixing and $\chi_2 \to \chi_1 + 3\gamma$ through an electron loop.
The width of $\chi_2\to\chi_1\nu\bar\nu$ is
\begin{equation}
\Gamma(\chi_2 \to \chi_1\nu\bar{\nu}) \simeq \frac{\alpha_{\rm EM}\, g_X^2\, \epsilon^2}{315\pi^2 c_W^2}\, \frac{\delta^9}{\mzp^4 m_Z^4}
\end{equation}
for $\delta \ll \mzp,\, \mchi$.

For the $\chi_2\to\chi_1+3\gamma$ decay, we integrate the Euler--Heisenberg width $\Gamma^{\rm EH}_{Z'^*\to3\gamma}(\sqrt{q^2})$~\cite{Pospelov:2008jk} over the invariant mass $\sqrt{q^2}$ of the off-shell $Z'$, which gives
\begin{equation}
\Gamma^{\rm EH}(\chi_2 \to \chi_1 + 3\gamma) \simeq \frac{17\,\alpha_{\rm EM}^4\, g_X^2\, c_W^2\, \epsilon^2}{1368241875\,\pi^5}\, \frac{\delta^{13}}{m_e^8\, \mzp^4}
\end{equation}
for $\delta \ll m_e$ and $\delta \ll \mzp,\, \mchi$, which coincides with the width of Ref.~\cite{Krnjaic:2025zjl}. 
As $\delta$ approaches $2m_e$, higher orders in $q^2/m_e^2$ matter~\cite{McDermott:2017qcg}. We rescale $\Gamma^{\rm EH}_{Z'^*\to3\gamma}(\sqrt{q^2})$ inside the integral over $0 \le q^2 \le \delta^2$ by the ratio of the full one-loop to the Euler--Heisenberg width tabulated in Ref.~\cite{McDermott:2017qcg}, keeping the full $Z'$ propagator and recoil kinematics, which enhances the width by a factor of seven at $\delta = 1\MeV$ and more as $\delta$ approaches $2m_e$.
On the low branch, $t_0/\tau_{\chi_2} \simeq 10$ at the best fit (Table~\ref{stab:benchmarks}), so that $f_2$ depends exponentially on $\Gamma_{\chi_2}$ there.

\Cref{sfig:ladder} illustrates the origin of the two branches discussed in the main text. As $\mzp$ increases, $f_2$ rises steeply and saturates near $1/2$, while $\sigma_{\chi p}$ falls as $\mzp^{-4}$, so that $N_{\roi}$ reaches one event twice.

\section{Other constraints}\label{ssec:constraints}
\begin{figure}[t]
    \centering
    \includegraphics[width=0.62\linewidth]{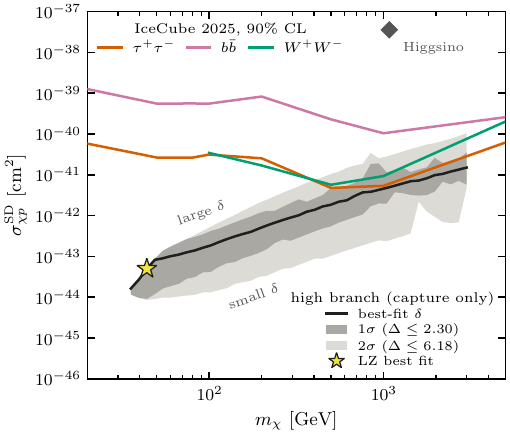}
    \caption{Effective solar cross section $\sigma^{\rm eff}_\odot$ along the high branch, for the best-fit $\delta$ at each $\mchi$ (line) and over the high-branch parts of the $1\sigma$ and $2\sigma$ regions of \cref{fig:allowed} (bands). The star marks the LZ best fit. The colored lines are the IceCube limits for single annihilation channels~\cite{IceCube:2025fcu}. The diamond marks a $1.08\TeV$ Higgsino~\cite{Pospelov:2026ewn}, with the capture rate of Ref.~\cite{DiMauro:2026dqp} for $\rho_\odot = 0.3\GeV/{\rm cm}^3$.}
    \label{sfig:solar}
\end{figure}

\dfsection{INTEGRAL limit}
We apply the INTEGRAL limit of Ref.~\cite{Krnjaic:2025zjl} to the total $\chi_2$ lifetime.
At $\delta = 1\MeV$, it reads $\tau_{\chi_2}\mchi/(2f_2) \geq 7.66\times10^{23}~{\rm s\;GeV}$, where $f_2 = 1/2$ for the equal mixture assumed there.
The decay flux scales with the $D$-factor, the line-of-sight integral of the DM density over the observed sky region.
Compared with their Navarro--Frenk--White (NFW) profile, our halo gives a $1.38$ times larger $D$-factor, which raises the lower bound on $\tau_{\chi_2}\mchi$ by the same factor.
Since $\Gamma_{3\gamma}$ grows steeply with $\delta$, a slightly smaller $\delta$ satisfies the limit at a smaller cost in $-2\ln\mathcal{L}$ than a lower LZ rate at $\delta = 1\MeV$.
The best INTEGRAL-allowed points thus lie at $\delta = 0.99\MeV$ (last two columns of Table~\ref{stab:benchmarks}).
The limit removes only 3 low-branch and 1 high-branch scan points from the $\Delta(-2\ln\mathcal{L}) \le 2.30$ regions, so we do not impose it in our figures and fits.

\dfsection{CMB limit}
To test the CMB constraint of the main text, we define the CMB ratio as the energy injected by the $\chi_2$ decays in units of the limit of Ref.~\cite{Slatyer:2016qyl}, with the $\chi_2$ fraction before its decays, $f_2^{\rm conv}$.
Since the CMB sensitivity varies with the energy and species of the injected particles~\cite{Slatyer:2016qyl}, this ratio depends on the photon spectrum. We therefore evaluate it in two ways.
First, we use the three-photon spectrum of Ref.~\cite{Krnjaic:2025zjl} and average the CMB sensitivity over its photon energies.
Second, as a conservative check, we place all the injected energy where the CMB is least sensitive, among photons and $e^\pm$ of any energy, which gives a bound independent of the spectrum.
This matters because the Euler--Heisenberg form of the spectrum is only approximate as $\delta$ approaches $2m_e$~\cite{Krnjaic:2025zjl}.

At the low-branch best fit, the ratio is $53.0$ with our spectrum (Table~\ref{stab:benchmarks}) and still $12.9$ with the least sensitive injection, so that this point is excluded in either case.
At the high-branch best fit, where $\chi_2$ lives much longer than the age of the Universe, the ratio is only $2.3\times10^{-3}$ with our spectrum and $4.8\times10^{-4}$ with the least sensitive injection.

\dfsection{Solar neutrinos}
DM captured in the Sun can annihilate there into high-energy neutrinos, which Super-Kamiokande, ANTARES, and IceCube search for~\cite{Super-Kamiokande:2015xms,ANTARES:2016xuh,IceCube:2025fcu}.
These searches limit the spin-dependent DM--proton cross section, assuming capture on solar hydrogen and equilibrium between capture and annihilation.
Solar capture of inelastic DM was recently studied in Refs.~\cite{DiMauro:2026dqp, Pospelov:2026ewn, Nguyen:2026lui}.
We compute our capture rate $C$ for single scattering on the nuclei of the standard solar model, with the present fraction $f_2$ and the Standard Halo Model~\cite{Baxter:2021pqo}.
We then compare it with the hydrogen capture rate $C_{\rm SD}$ that these searches assume for a spin-dependent cross section of $10^{-40}~{\rm cm}^2$~\cite{IceCube:2025fcu}, through $\sigma^{\rm eff}_\odot \equiv 10^{-40}~{\rm cm}^2 \times C/C_{\rm SD}$.
At the high-branch LZ best fit, $\sigma^{\rm eff}_\odot = 5.0\times10^{-44}~{\rm cm}^2$ (Table~\ref{stab:benchmarks}), about 600 times below the IceCube $\tau^+\tau^-$ limit at this mass.
The comparison is conservative.
Treating every annihilation as a $\tau^+\tau^-$ pair likely overestimates the neutrino yield of the $Z'$ cascades.
Only for $\mchi \gtrsim 400\GeV$ does part of the high-branch region reach the $\tau^+\tau^-$ and $W^+W^-$ limits (\cref{sfig:solar}).
On the low branch, $\sigma^{\rm eff}_\odot = 5.4\times10^{-44}~{\rm cm}^2$ is similar, but the muons and mesons from $Z'$ decays stop in the solar core before they decay, so their neutrinos fall below the search thresholds.

\section{Galactic Center excess}\label{ssec:gce}
We compute the prompt $\gamma$-ray intensity from DM annihilation in the inner Galaxy and compare it with three GCE spectra.
For the present mixture of $\chi_1$ and $\chi_2$, the intensity is 
\begin{equation}
E^2 d\Phi/dE\,d\Omega = \bar J/(8\pi\mchi^2)\sum_{ij} w_{ij}\langle\sigma v\rangle_{ij}E^2 dN_{ij}/dE,
\end{equation} with $w_{11} = f_1^2$, $w_{12} = 2f_1f_2$, and $w_{22} = f_2^2$, so that $\langle\sigma v\rangle_{\rm eff} = \sum_{ij} w_{ij}\langle\sigma v\rangle_{ij}$.
Here $\bar J$ is the $J$-factor of our halo, the line-of-sight integral of the squared DM density averaged over the data region.
The prompt photon spectra $dN_{ij}/dE$, including the $Z'$ and $\phi$ cascades, come from micrOMEGAs~\cite{Belanger:2026asz}.
For the low-branch points, whose $Z'$ decays into leptons and light mesons, micrOMEGAs gives no photons from hadronic $Z'$ decays for $\mzp < 1.2\GeV$.
We therefore compute their spectrum with the exclusive hadronic channels of Hazma~\cite{Coogan:2019qpu,Coogan:2022cdd,Plehn:2019jeo}, which are valid for $\mzp \lesssim 1.5\GeV$, and boost it to the annihilation frame. The $\phi$ channels play no role here, since $\chi_1\chi_2 \to Z'\phi$ is suppressed by the small $f_2$, $1.5\times10^{-6}$ at the best fit, and $\chi_i\chi_i \to \phi\phi$ by $\delta/\mzp$.

We use three GCE spectra, the southern-sky spectrum of Ref.~\cite{Cholis:2021rpp} ($|l| \le 20^\circ$, $-20^\circ \le b \le -2^\circ$) as the main one, the full $40^\circ\times40^\circ$ window of the same analysis, and the spectrum of Di Mauro~\cite{DiMauro:2026fnp}.
In the north, the bright diffuse emission from interstellar gas makes even small modeling errors significant, so that every model considered in Ref.~\cite{Cholis:2021rpp} fails in the full window.
Its authors therefore recommend fitting the southern sky alone.
We quantify the agreement with the data by a $\chi^2$ that includes the covariance $\Sigma$ between energy bins.
For the spectra of Ref.~\cite{Cholis:2021rpp}, $\Sigma$ combines uncorrelated statistical errors with correlated systematic errors. For the southern sky, whose statistical errors are not published, we take them as the half-width of the $1\sigma$ band.
For the latter, we use the published systematic covariance matrix, reduced to its three leading principal components and scaled by $0.45^2$ for the southern sky, as prescribed in Ref.~\cite{Cholis:2021rpp}.
For the spectrum of Di Mauro~\cite{DiMauro:2026fnp}, the statistical errors and the spread over the different Galactic diffuse-emission models are given.
We take half this spread as the systematic error.
Since its correlation between bins is not given, we consider uncorrelated and fully correlated systematic errors as two limits.

\begin{figure}[t]
    \centering
    \includegraphics[width=\linewidth]{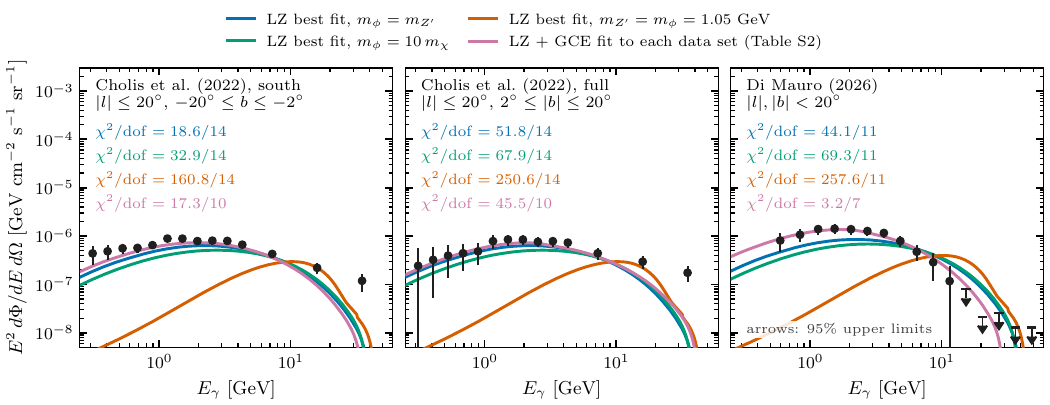}
    \caption{$\gamma$-ray intensity with $A = 1$ on the southern-sky and full-window spectra of Ref.~\cite{Cholis:2021rpp} and the spectrum of Di Mauro~\cite{DiMauro:2026fnp} (arrows: upper limits). Blue: high-branch LZ best fit ($\mzp = \mphi = 25.1\GeV$). Green: LZ best fit with $\mphi = 10\,\mchi$ ($\mzp = 29.0\GeV$). Orange: low-branch LZ best fit ($\mzp = \mphi = 1.05\GeV$). Purple: joint LZ + GCE fit to each data set. The error bars are the square roots of the covariance diagonals, and the $\chi^2/{\rm dof}$ values are those of Table~\ref{stab:gce}.}
    \label{sfig:gce-flux}
\end{figure}

On the low branch, the spectrum of the LZ best fit peaks near $10\GeV$ and gives $\chi^2/{\rm dof} = 160.8/14$, $250.6/14$, and $257.6/11$ on the three data sets, worse than no signal, which gives $130.7$, $163.8$, and $141.2$ (\cref{fig:lz-best-fit}, \cref{sfig:gce-flux}). With a free normalization, the southern sky gives $A = 0.16 \pm 0.15$ (Table~\ref{stab:benchmarks}), consistent with no signal.
The low branch is therefore disfavored by the GCE data rather than merely unable to explain them.

Without any fit to the GCE, the high-branch LZ best fit gives $\chi^2/{\rm dof}=18.6/14$ ($p = 0.18$) for the southern sky, $51.8/14$ for the full window, and $44.1/11$ for Di Mauro~\cite{DiMauro:2026fnp} (\cref{sfig:gce-flux}). 
A free normalization gives $A = 1.06 \pm 0.10$ for the southern sky, consistent with the relic cross section.

\begin{figure}[t]
    \centering
    \includegraphics[width=0.49\linewidth]{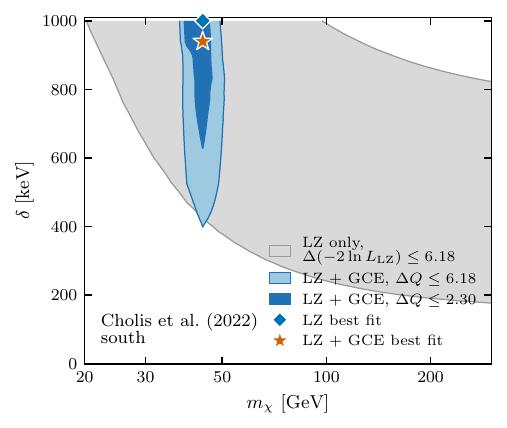}
    \caption{Regions of the $(\mchi,\delta)$ plane favored by the LZ event alone (gray, $\Delta(-2\ln\mathcal{L}_{\rm LZ}) \le 6.18$) and together with the southern-sky GCE spectrum of Ref.~\cite{Cholis:2021rpp} (light and dark blue, $\Delta Q \le 6.18$ and $2.30$).
At each point, $Q$ is minimized over $\mzp$ and $\mphi/\mzp \in \{1/2, 1, 2\}$, with $A = 1$.
The diamond marks the LZ best fit and the star the best fit of this three-ratio scan, which has $\mphi = \mzp$.}
    \label{sfig:gce-region}
\end{figure}

For the joint fits of the main text, we minimize
\begin{equation}
Q = -2\ln\mathcal{L}_{\rm LZ} + \chi^2_{\rm GCE},
\label{seq:joint-q}
\end{equation}
with $\mathcal{L}_{\rm LZ}$ the LZ likelihood of Eq.~\eqref{eq:likelihood}.
With $\mphi = \mzp$, the southern-sky best fit lies at $(43.9\GeV, 0.94\MeV, 30.4\GeV)$, with $\chi^2/{\rm dof} = 17.2/11$ at $\Delta(-2\ln\mathcal{L}_{\rm LZ}) = 0.41$. Profiling $Q$ in $\mchi$ gives $\Delta Q \le 1$ for $\mchi = 39.5$--$43.9\GeV$, the interval quoted in the main text. Unlike the LZ region, which extends to TeV masses, the joint region is confined to the light end of the LZ valley (\cref{sfig:gce-region}).

Letting $\mphi$ vary as well, the four-parameter fit of Table~\ref{stab:gce} lies at $(39.5\GeV, 1\MeV, 27.6\GeV)$ with $\mphi/\mzp = 0.71$ and has $\chi^2/{\rm dof} = 17.3/10$ at $\Delta(-2\ln\mathcal{L}_{\rm LZ}) = 0.23$ ($16.8/9$ at $0.08$ with $A = 1.12$).
In the full window, the best joint fit still has $\chi^2/{\rm dof} = 45.5/10$ at $\Delta(-2\ln\mathcal{L}_{\rm LZ}) = 4.05$.
For the spectrum of Di Mauro, it has $3.2/7$ at $\Delta(-2\ln\mathcal{L}_{\rm LZ}) = 1.24$ with uncorrelated systematic errors, but $\chi^2 = 50.9$ with fully correlated ones and a free normalization.

To constrain $\mphi$, we profile $Q$ over $(\mchi, \delta, \mzp)$ at fixed $\mphi/\mzp$, relative to the four-parameter best fit.
$\Delta Q \le 1$ holds for $0.35 \le \mphi/\mzp \le 1.4$ ($0.5$--$2$ with a free normalization), where $\chi_1\chi_2 \to Z'\phi$ is open. Profiling $Q$ in $\mphi$ itself gives $\Delta Q \le 1$ for $\mphi \approx 16$--$39\GeV$ ($12.5$--$55\GeV$ for $\Delta Q \le 4$).
At $\mphi/\mzp = 2.8$, where this channel is closed at a perturbative $\lambda_\Phi \simeq 0.32$, $\Delta Q = 6.17$.

\Cref{sfig:gce-flux} compares our spectra with the three data sets, whose differences reflect the modeling of the interstellar emission.
For all three, the low branch fits worse than no signal, and closing $\chi_1\chi_2 \to Z'\phi$ worsens the high-branch fit, so that the dark Higgs boson remains needed whichever spectrum is used (Table~\ref{stab:benchmarks}).

\begin{table}[t]
\centering
\caption{$\chi^2$ of the predicted $\gamma$-ray spectra for three GCE spectra.
The first three rows are the LZ best fits of Table~\ref{stab:benchmarks} and \cref{fig:lz-best-fit}, the third with $\mphi = 10\,\mchi$, without any fit to the GCE, for the common halo, with $\chi^2$ given per number of bins.
The joint rows minimize $Q$ with the LZ term for the common halo, over $(\mchi, \delta, \mzp, \mphi)$, with $A = 1$ except in the southern-sky column of the free-$A$ row ($A = 1.12$). For the spectrum used in each joint fit, the denominator is the number of bins minus the four minimized parameters (five with $A$), and these values are descriptive rather than goodness-of-fit measures.
$\Delta(-2\ln\mathcal{L}_{\rm LZ})$ is measured from the LZ best fit.
For the southern sky, whose statistical errors are not published, we use the half-width of the $1\sigma$ band of the spectrum and scale the systematic covariance by $0.45^2$, following Ref.~\cite{Cholis:2021rpp}.
For the low-branch point, whose $Z'$ decays into leptons and light mesons, the photon spectrum includes the hadronic $Z'$ decays computed with Hazma (see text).}
\label{stab:gce}
\footnotesize
\setlength{\tabcolsep}{1.8pt}
\begin{tabular}{@{}lccccccccccc@{}}
\toprule
 & $\mchi$ & $\delta$ & $\mzp$ & $\mphi$ & $g_X$ & $f_2$ & $\langle\sigma v\rangle_{\rm eff}$ & \multicolumn{2}{c}{Ref.~\cite{Cholis:2021rpp}} & Di Mauro~\cite{DiMauro:2026fnp} & $\Delta(-2\ln\mathcal{L}_{\rm LZ})$ \\
 & [GeV] & [keV] & [GeV] & [GeV] & & & [$10^{-26}~{\rm cm^3/s}$] & south & full & & \\
\midrule
LZ best fit, low & 43.9 & 1000 & 1.05 & 1.05 & 0.1162 & $1.5\times10^{-6}$ & 2.20 & 160.8/14 & 250.6/14 & 257.6/11 & 0.00 \\
LZ best fit, high & 43.9 & 1000 & 25.1 & 25.1 & 0.1112 & 0.497 & 2.32 & 18.6/14 & 51.8/14 & 44.1/11 & 0.00 \\
High, $\mphi = 10\,\mchi$ & 43.9 & 1000 & 29.0 & 439.1 & 0.1488 & 0.496 & 2.25 & 32.9/14 & 67.9/14 & 69.3/11 & 0.00 \\
\multicolumn{12}{@{}l}{LZ + GCE, minimized:} \\
\quad south, 4D & 39.5 & 1000 & 27.6 & 19.5 & 0.1015 & 0.498 & 2.27 & 17.3/10 & 55.4/14 & 23.5/11 & 0.23 \\
\quad south, 4D, free $A$ & 43.9 & 1000 & 27.6 & 39.0 & 0.1157 & 0.497 & 2.25 & 16.8/9 & 52.4/14 & 31.2/11 & 0.08 \\
\quad full, 4D & 43.9 & 800 & 53.2 & 13.3 & 0.0869 & 0.500 & 2.38 & 15.3/14 & 45.5/10 & 63.5/11 & 4.05 \\
\quad Di Mauro, 4D & 32.0 & 1000 & 29.3 & 14.7 & 0.0861 & 0.499 & 2.26 & 53.5/14 & 112.2/14 & 3.2/7 & 1.24 \\
\bottomrule
\end{tabular}
\end{table}

\putbib[LZ2026,LZ2026-aux]
\end{bibunit}
\end{document}